\documentclass[pre,letterpaper,aps,twocolumn]{revtex4}
\usepackage{graphicx}
\usepackage{amsmath}
\usepackage[active]{srcltx}
\usepackage{amsfonts}
\usepackage[latin1]{inputenc}
\graphicspath{{FIGURES/}}

\begin{document}
\title{Evolving complex networks with conserved 
       clique distributions}

\author{Gregor Kaczor and Claudius Gros}
\affiliation{Institute for Theoretical Physics, 
             Johann Wolfgang Goethe University, 
             Frankfurt am Main, Germany}
\date{\today}

\begin{abstract}
We propose and study a hierarchical algorithm 
to generate graphs having a predetermined distribution 
of cliques, the fully connected subgraphs. The
construction mechanism may be either random or
incorporate preferential attachment.
We evaluate the statistical properties of 
the graphs generated, such as the
degree distribution and network diameters,
and compare them to some real-world graphs.
\end{abstract}
\maketitle

\section{Introduction}

The structural and statistical properties 
of networks have been studied intensively 
over the last decade \cite{Albert02,Dorogovtsev02},
due to their ubiquitous importance
in technology, different realms of life and
complex system theory in general \cite{Gros07}.
With time it was realized that
the topological properties of
real-world networks often transcend
the universality class of both the
straightforward, all-random 
Erd\"os-R\'enyi graph \cite{Erdos59},
as well as that of
random networks with arbitrary 
degree distributions \cite{Newman01}.

Many real-world networks have a well defined
community  structure \cite{Palla05}. A
community is, loosely speaking, a 
subgraph which has an intra-subgraph link density
which is substantial above the average
link-density of the whole network. The community with
link density equal to one is denoted
in graph theory as a `clique'. A clique is a fully 
interconnected subgraph, the smallest clique
having just two vertices.

A clique is also a specific realization of
a graph motif, i.e.\ of subgraphs with definite 
topologies \cite{Milo02,Vazquez04},
and of $k$-cores, {\it viz} subgraphs with at least
$k$ interconnections \cite{Dorogovtsev06}.
In a related work Derenyi {\it et al.} have
introduced the notion of clique percolation 
in the context of overlapping graph communities

\cite{Derenyi05}. For scale free graphs, having
 a degree distribution $p_k\sim k^{-\gamma}$,
the second moment $\langle k^2\rangle$ diverges for the
important case $2<\gamma<3$ and finite numbers
of cliques of arbitrary size emerge \cite{Bianconi06}.

For any graph one can define a characteristic
clique distribution $P_C(S)$, {\it viz} the probability
for a clique of size $S$ to occur.
A loopless graph, exclusively has, 
cliques of size two with
$P_C(S)=\delta_{S,2}$ and the number of
3-site cliques is related to the 
standard clustering coefficient 
\cite{Albert02,Dorogovtsev02}. The
clustering coefficient $C$ is a normalized
measure for the occurrence of 3-site loops,
with every 3-site loop being part of at
least one clique of size $S\ge3$.

It is therefore of interest to investigate
the clique distribution of real-world graphs 
and to consider the problem of constructing
graphs with specific clique distributions.

\section{Algorithm}

\begin{figure}[t]
\centerline{
\includegraphics[width=0.40\textwidth]{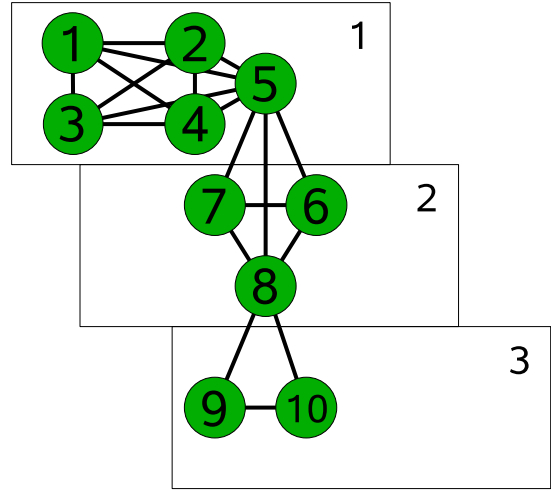}
           }
\caption{Illustration of a clique-conserving
algorithm generating a connected graph out of a 
given set of cliques. Starting with a 5-site 
clique (1,2,3,4,5) in step one, a 4-site clique
(5,6,7,8) and a 3-site clique (8,9,10) are 
added in step two and three via a single common 
vertex.}
\label{lattice_of_cliques} 
\end{figure}

We consider a given set of cliques
$C_1,\dots, C_M$ containing $S_i=S(C_i)$
sites each, an instantiation of a certain
clique distribution $P_C(S)$. We presume the clique-set
to be monotonically ordered,
\begin{equation}
S_i \ \ge S_{i+1},\qquad i=1,\dots,M-1~,
\label{eq:clique_ordering}
\end{equation}
as illustrated in Fig.\ \ref{lattice_of_cliques}.
We study the task to generate recursively
a dense and connected graph out of the 
$M$ cliques $\{C_i\}$ in such a
way that the final graph has exactly the
same distribution $P_c(S)$
of fully connected subgraphs,
{\it viz} of cliques. In 
Fig.\ \ref{lattice_of_cliques} we
illustrate the simplest procedure
for solving this task, by
concatenating the cliques
$C_1,\ C_2,\ ..$ via a single
common vertex between two consecutive 
cliques.

Let us shortly digress and consider what would
have happened if we had used sites 4 and 7, 
together with a new site 9 to attach 
the $S_3=3$ clique in the third step for the
case illustrated in Fig.\ \ref{lattice_of_cliques}.
In this case sites 4 and 7 would be connected
and a spurious 3-site clique, namely (4,5,7),
would have been generated. A thoughtless attachment
of cliques in general therefore generates
spurious additional cliques, resulting in an
uncontrolled clique distribution for the final
graph.

\subsection{Hierarchical algorithm}

\begin{figure}[t]
\centerline{
\includegraphics[width=0.40\textwidth]{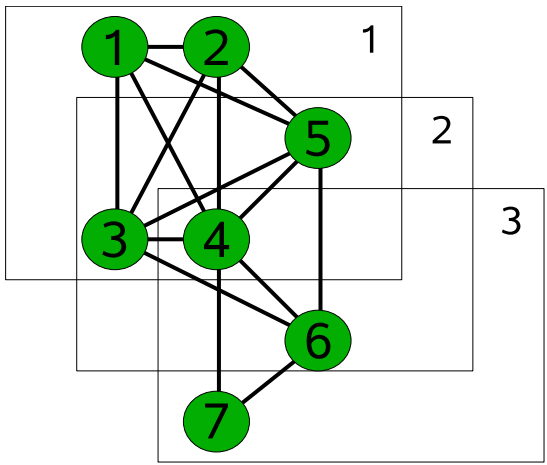}
           }
\caption{Illustration of the dense hierarchical algorithm for
generating dense graphs out of a given set of cliques.
Starting with the largest clique (1,2,3,4,5), 
here of size $S_1=5$, in step 1, 
cliques of size $S_i$ ($i=2,3,\dots$) are added 
consecutively step by step by adding one 
additional vertex at each step and using
$S_i-1$ vertices from a previously added clique.
The second clique is here (3,4,5,6) and the third
clique (4,6,7). Both random and preferential 
attachment may be used.
         }
\label{fig_algorithm_steps} 
\end{figure}

In general one can join two cliques of sizes $S_1$ and $S_2$
via common vertices. The minimal number of common
vertices is one, the maximal is
\begin{equation}
\label{eq:min_sites}
min(S_1,S_2)-1~.
\end{equation}
Using more common sites, namely $min(S_1,S_2)$,
would result in the destruction of the smaller
clique. We can then formulate a class of 
hierarchical algorithms conserving a given,
arbitrary but ordered, via (\ref{eq:clique_ordering}),
initial clique distribution:
\begin{description}
\item [[1]]
      At step $m=1,...,M$ one adds the clique $C_m$
      with $S_m=S(C_m)$ sites. One starts by selecting
      a number $\tilde S_m\in[1,S_m-1]$. Here we will
      mostly concentrate on the case $\tilde S_m = S_m-1$.
\item [[2]]
      Next one selects recursively $\tilde S_m$
      mutually interconnected vertices out of the
      graph segment
      constructed in the previous $m-1$ steps.
      The new clique is then added by mutually 
      connecting $S_m-\tilde S_m$ new sites
      among themselves and with the $\tilde S_m$
      selected sites of the existing graph segment.

\end{description}
We call the choice $\tilde S_m=S_m-1$ 
the `dense hierarchical algorithm';
it is illustrated in
Fig.\ \ref{fig_algorithm_steps}. Here
we will study exclusively the dense
algorithm, which results in quite 
dense networks. The opposite limit, namely
the case $\tilde S_m=1$ in step [1] of the
hierarchical algorithm, is illustrated in
Fig.\ \ref{lattice_of_cliques}. 

Starting with $M$ cliques
the dense hierarchical algorithm
generates a network containing 
$N$ sites in its final state,
with 
\begin{equation}
N\ =\ S_1\,+\,(M-1)~,
\label{eq_N_S_M}
\end{equation}
with $S_1$ being the size of the
starting clique, which is also the largest.
This is so, because exactly one new vertex is
added at each of the $(M-1)$ steps.

\subsection{Random vs.\ preferential attachment}

The selection of the $\tilde S_m$ vertices in
step [2] can be done either randomly, by
preferential attachment or other rules.
When considering preferential attachment 
we first select a single vertex $i$ 
with an attachment probability $\Pi(k_i)$
proportional to the vertex-degree $k_i$,
\begin{equation}
 \Pi(k_i)\ =\  \frac{k_i}{\sum_{j}^{J} (k_j)}
\label{eq_preferential_attachment}
\end{equation}
(linear preferential attachment). 
We then select recursively $\tilde S_m-1$
vertices out of the neighbors of $i$
via preferential attachment. The set of
possible vertices is given, at every step
of this recursive selection process, by the
set of vertices linked to all sites
previously selected. Note that the
ordering (\ref{eq:clique_ordering})
of the initial clique distribution
is a precondition for the hierarchical
algorithm to function.

\subsection{Decimation algorithm}

For further reference we shortly mention a
second clique-conserving algorithm for 
network construction via vertex decimation.
Starting with an initial network of
$M$ unconnected cliques $C_1,...,C_M$ one
selects pairs of unconnected vertices 
either randomly or via preferential 
attachment. One then attempts a 
decimation by merging the two selected
vertices into a single vertex. 
One then calculates the clique distribution 
of the new network which has one less site.
If the new clique distribution is identical
to the original distribution the decimation is 
accepted, or else it is rejected.

\begin{figure}[t]
\centerline{
\includegraphics[width=0.45\textwidth]{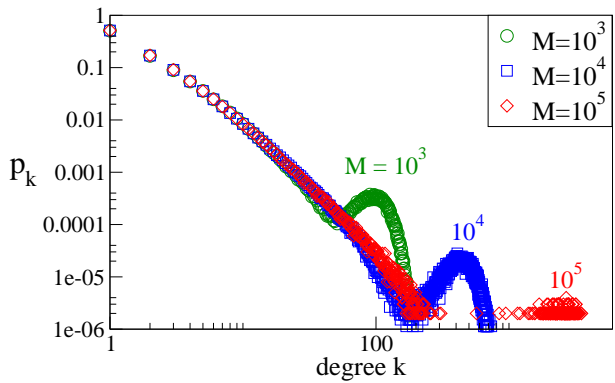}
           }
\caption{The degree distributions $p_k$ for
graphs with scale-free clique distribution, compare
Eq.\ (\ref{eq_scale_free_P_C}), and an exponent
$\alpha=2.6$.
Blue squares are for a system of $M\approx10^5$ 
cliques, green stars and red diamonds show systems for
$10^4$ and $10^3$ cliques respectively. The data
is obtained by averaging over $1000/263/19$
realizations for $P_C(S)$ for clique-numbers
$M$ equal to $10^3/10^4/10^5$.}
\label{same_alpha26_diff_sizes} 
\end{figure}

\section{Simulation results}

We have studied the properties of the hierarchical
clique-conserving graph-generation algorithm
extensively using numerical results, evaluating
their respective statistical properties and 
comparing them to some selected real-world graph.

\subsection{Initial clique distribution}

The hierarchical graph generation algorithm, conserves
per construction the initial clique distribution
$P_C(S)$. We have studied two cases. In
Sect.\ \ref{sect_compare_real_world} we will discuss
the results obtained by using the
measured clique distribution of real-world networks
for $P_C(S)$. Here we will concentrate on some
of the model clique distributions, in particular of
scale-free form
\begin{equation}
P_C(S)\ \sim\ \left({1\over S}\right)^\alpha,
\qquad\quad \alpha>2~.
\label{eq_scale_free_P_C}
\end{equation}
We performed simulations for various exponents $\alpha$,
and scale-free clique-distributions
containing a total number $M$ of cliques.
For the simulations a cut-off $S_1$
needs to be chosen for the scale-free 
distribution (\ref{eq_scale_free_P_C}), 
i.e.\ the maximal clique-size $S_1$.
The expected number $N_{S_1}(S)$ of
cliques is then
\begin{equation}
\label{eq_max_clique_size}
 N_{S_1}(S)\ =\ \left(1\over S\right)^{\alpha} 
    \frac{M}{\sum_{S'=1}^{S_1} (1/S')^{\alpha}}~,
\end{equation}
where $M$ is the total number of cliques.
We selected $S_1$ by the condition
\begin{equation}
N_{S_1}(S_1)\ >\ 1, \qquad\quad N_{S_1}(S_1+1)\ <\ 1~,
\label{S_1_selection}
\end{equation}
{\it viz} that there is at least one clique
of size $S_1$ present on the average. 
We compared results obtained for
$M$ ranging typical from $10^3-10^5$, in order
to extract scaling properties in the large-network
limit. In order to extract reliable statistical
properties the results were averaged over
$N_{real}$ different random realizations.

When selecting the value $S_1$ for the maximal 
clique size one discards all cliques with sizes
$S>S_1$. This is admissible when the percentage
of discarded cliques is small. With the criteria
(\ref{S_1_selection}) the percentage of discarded
cliques vanishes in the thermodynamic limit $M\to\infty$.
For the system of order $10^4,\ 10^5$, the 
percentage of discarded is well below $1\%$.

\begin{table}[b]
\caption{Statistical properties of graphs 
(compare Fig.\ \ref{same_alpha26_diff_sizes})
containing $M \approx 10^3,\ 10^4,\ 10^5$ cliques generated by the 
hierarchical algorithm with preferential attachment,
using a scale-free clique distribution
(\ref{eq_scale_free_P_C}), with an exponent $\alpha=2.6$.
$C$ is the clustering coefficient, $\ell$ 
the average path length, $\langle\kappa\rangle$ the
average degree, $D$ the network diameter, $d$ 
the link density and $N$ the total number 
of vertices. $m$ is the slope of the degree
distribution $p_k$ measured for $k\in[10,40]$ for 
$M \approx 10^3$ and  $k\in[10,100]$ for 
$M \approx 10^4,\ 10^5$.}
\begin{tabular}{cccccccc}
\hline \hline
$M$ &  $C$ & $\ell$ &$D$&  $\langle \kappa \rangle$ & $d$ & $N$ & $m$\\
\hline
$986$   & 0.34 & 3.2 & 7.5 & 5.1 & 0.00508  & 1007   & -2.6 \\
$9979$  & 0.36 & 3.3 & 8.8 & 5.7 & 0.00056  & 10032  & -2.7 \\
$99999$ & 0.37 & 3.4 & 9.8 & 5.8 & 0.000058 & 100096 & -2.4 \\
\hline \hline
\end{tabular}
\label{table_comparison_diff_sizes}
\end{table}

\begin{figure}[t]
\centerline{
\includegraphics[width=0.45\textwidth] {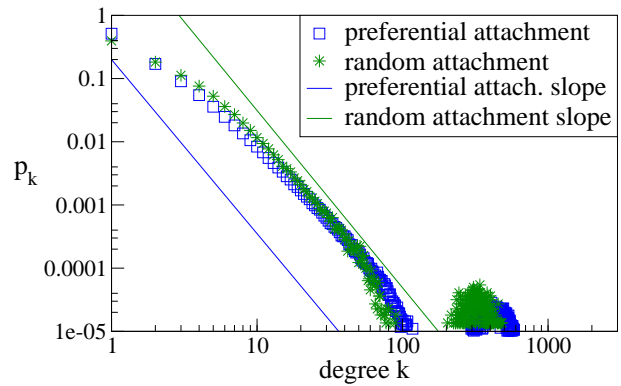}
           }
\caption{The degree distribution $p_k$ for graphs having
a scale-free clique distribution with $\alpha=2.6$ and $M\approx10^4$ 
cliques. Shown are results both for random and 
preferential attachment with lines indicating
the respective slopes $-2.7$ (random) and 
$-2.5$ (preferential).}
\label{com_rnd_prefAtt} 
\end{figure}

\subsection{System-size analysis}

In Fig.\ \ref{same_alpha26_diff_sizes} we present
the degree distribution $p_k$ for graphs with
a scale-free clique distribution (\ref{eq_scale_free_P_C})
and an exponent $\alpha=2.6$, generated through
the hierarchical algorithm with preferential attachment.
The degree distribution results from averaging
$N_{real}=1000,\ 263,\ 10 $ realizations for
clique distributions containing $M\approx10^3,\ 10^4,\ 10^5$
cliques. We note that the degree distribution
approaches a well defined curve for the thermodynamic
limit $M\to\infty$.

The degree distributions shown in 
Fig.\ \ref{same_alpha26_diff_sizes} have bumps at
high degrees for finite numbers of cliques $M$. 
This is due to the fact that the algorithm 
starts by incorporating the large cliques first 
so that vertices with an high initial degree 
see it further increased via the preferential
attachment during the construction process.
This effect vanishes in the thermodynamical limit as the
probability of a given vertex to be chosen as a part of a new
clique decreases with system size.

The statistical analysis of the networks presented
in Fig.\ \ref{same_alpha26_diff_sizes} are given
in Table \ref{table_comparison_diff_sizes},
the number of cliques $M$ and
the number of vertices, $N$ obey
the relation (\ref{eq_N_S_M}) valid
for the dense hierarchical algorithm.
The resulting degree distribution, approaches
within the numerical errors, a scale-free
functional dependence with an exponent $|m|$
approximately given by the exponent
$\alpha=2.6$ of the conserved clique
distribution $P_C(S)$.

In Fig.\ \ref{com_rnd_prefAtt} we compare
the degree distribution between 
construction rules with preferential and
random attachment respectively.
The difference is quite small in the region of small
to intermediate degrees $k$, where the finite-size
corrections are minor, the reason being
the algorithmic restriction, that only common 
neighbors of the already processed vertices can 
be used to construct a clique iteratively.
This restriction decreases the number of
vertices available for the preferential attachment
and results in a similar degree distribution, which
is however slightly different from the ideal scale
free line.

\begin{figure}[t]
\begin{center}
\includegraphics[width=0.45\textwidth] {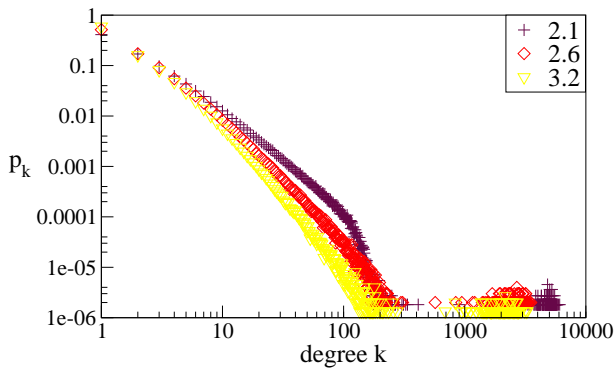}
\end{center}
\caption{\label{comparison_differentalphas} 
The degree distribution for three scale free 
clique distributions with exponents
$\alpha_1=2.1$ (maroon plus, $11$ simulation runs), 
$\alpha_2=2.6$ (red diamonds, $26$ simulation runs), 
$\alpha_3=3.2$ (yellow triangles, $15$ simulation runs),
for $M\approx10^5$ cliques.}
\end{figure}

\subsection{Dependency on the scaling exponent} 

We have studied the properties of the graphs generated
by the hierarchical algorithm for scale-free clique 
distributions $P_C(S)$ and several scaling exponents 
$\alpha$. We have analyzed the corresponding
graphs as a function of clique-numbers $M\approx10^3,\ 10^4,\ 10^5$,
averaging over several clique-distribution realizations. 
The resulting degree distributions are shown in
Fig.\ \ref{comparison_differentalphas} 
for the case  $M\approx10^5$, the corresponding
statistical analysis in Table \ref{table_comparison_diff_alpha105}.
In order to estimate the finite-size corrections we 
present in Table \ref{table_comparison_diff_alpha104}
the corresponding results for $M\approx10^4$. We note,
in particular, a good agreement in the estimates for
the scaling exponent $|m|$ of the resulting 
degree distribution. 

Interestingly enough, the exponent $|m|$ for the degree
distribution of the graph generated by the 
hierarchical algorithm with preferential attachment
saturates at $\approx3.1$, close to the value $3$ expected
for the standard preferential attachment algorithm
\cite{Albert02}. When $\alpha<3$ the large tail
of the degree distribution stemming directly from 
the clique distribution dominates the resulting
exponent $|m|$ for the degree distribution, but
fails to do so for $\alpha>3$, when the
preferential attachment mechanism dominates the
generation of the fat tail.

Next, we comment on the size of the network diameter
$D$ of the generated graphs.
With increasing $\alpha$ we observe an increasing 
average path length $\ell$ and an increasing average 
diameter $D$ while the clustering coefficient $C$ 
decreases. The network diameter is intuitively
affected by the number of low-degree vertices.
A larger number of low-degree vertices for
degree distributions of identical functional 
dependences, generally results in a bigger network diameter.
Alternatively one may consider the number of
trivial cliques, namely those with size $S=2$, {\it viz}
edges not forming part of any larger clique.
They tend to connect to low-degree vertices, since two
connected high-degree vertices would have a higher probability 
to belong to cliques of size 3 or larger.

\begin{table}[bt]
\caption{Statistical properties of graphs 
(compare Fig.\ \ref{comparison_differentalphas})
containing $M \approx 10^5$ cliques generated by the 
hierarchical algorithm with preferential attachment.
$\alpha$ denotes the scaling exponent for the
clique distribution $P_c(S)$,
$C$ the clustering coefficient, $\ell$ 
the average path length, $\langle\kappa\rangle$ 
the average degree, $D$ is the network diameter, $d$ 
is the link density and $N$ the total number 
of vertices. $m$ is the slope of the degree
distribution $p_k$ measured for $k\in[10,100]$.}
\begin{tabular}{cccccccc}
\hline \hline
$\alpha$ &  $C$ & $\ell$ &$D$&  $\langle \kappa \rangle$ & $d$ & $N$ & $m$\\
\hline
2.1 & 0.51 & 3.2 &  9.2 & 9.9 & 0.000098 & 100099 & -2.1 \\
2.6 & 0.36 & 3.4 &  9.8 & 5.8 & 0.000058 & 100096 & -2.4 \\
3.2 & 0.23 & 3.7 & 11.3 & 3.8 & 0.000038 & 100042 & -3.1 \\
4.2 & 0.10 & 4.1 & 13.4 & 2.8 & 0.000027 & 100020 & -3.2 \\
\hline \hline
\end{tabular}
\label{table_comparison_diff_alpha105}
\end{table}

In order to examine the influence of these trivial cliques
on the network diameter we have eliminated, from the
graph generated by the hierarchical algorithm with\
$M\approx10^4$ and $\alpha=2.1,\ 2.6,\ 3.2,\ 4.2$
all cliques of size $S=2$.
The statistical properties of the resulting graph 
are given in
Table \ref{table_comparison_diff_alpha104}.
The network diameter $\ell$ decreases substantially 
and the clustering $C$ increases.
We note that the scaling exponent $m$ for the
degree distribution remains unaffected, as it
depends on the vertices with large degrees only.
This result is nevertheless somewhat surprising,
in view of dramatic reduction in the number of
vertices $N$ resulting from the decimation of
all trivial cliques. 

\begin{table*}[bth]
\caption{Left table: Statistical properties 
of graphs with $M\approx 10^4$ cliques and
various scaling exponents $\alpha$ 
for the clique distribution $P_C(S)$.
$C$ is the clustering coefficient, $\ell$ 
the average path length, $\langle\kappa\rangle$ 
the average degree, $D$ the network diameter, 
$d$ the link density, $N$ the total number 
of vertices and $m$ the slope measured 
between degree 10 and 60. The degree distributions
result from averaging $N_{real}=86, 263, 866, 306$ 
realizations for clique distributions having 
$\alpha=2.1,\ 2.6,\ 3.2,\ 4.2$. \newline
Right table: The same data as for the left table, but
with all cliques of degree $S=2$ removed from the graphs.}

\centerline{\hfill
\begin{tabular}{ccccccccc}
\hline \hline
$\alpha$ &  $C$ & $\ell$ &$D$&  $\langle \kappa \rangle$ & $d$ & N & $m$\\
\hline
2.1 & 0.51 & 3.1 & 7.5  & 10.5 & 0.00104 & 10093 & -2.0 \\
2.6 & 0.36 & 3.4 & 8.8  & 5.6  & 0.00056 & 10032 & -2.5 \\
3.2 & 0.23 & 3.7 & 10.0 & 3.8  & 0.00038 & 10017 & -2.9 \\
4.2 & 0.10 & 4.1 & 11.8 & 2.7  & 0.00027 & 10007 & -3.1 \\
\hline \hline
\end{tabular}
\hfill
\begin{tabular}{ccccccccc}
\hline \hline
$\alpha$ &  $C$ & $\ell$ &$D$&  $\langle \kappa \rangle$ & $d$ & N & $m$\\
\hline
2.1 & 0.94 & 2.73 & 4.1 & 16.7  & 0.0028  & 5885  & -2.0 \\
2.6 & 0.92 & 2.77 & 4.5 & 10.0  & 0.0021  & 4625  & -2.5 \\
3.2 & 0.90 & 2.77 & 4.9 & 7.2   & 0.00206 & 3491  & -3.0 \\
4.2 & 0.97 & 2.74 & 5.0 & 5.5   & 0.0024  & 2207  & -3.0 \\
\hline \hline
\end{tabular}
   \hfill}
\label{table_comparison_diff_alpha104}
\end{table*}

\section{Comparison with real world data}
\label{sect_compare_real_world}

We have evaluated the clique distributions $P_C(S)$
for two real-world networks, a
protein-protein interaction network
\cite{Mathivanan06} and a WWW-graph 
\cite{Barabasi99}. We then have used
the resulting clique distributions $P_C(S)$,
as the starting point for the hierarchical
algorithm with preferential attachment and
compared the hence generated graphs
with the properties of the
original real-world networks.

\begin{figure*}[htbp]
  \centerline{
    \mbox{\includegraphics[width=0.45\textwidth] {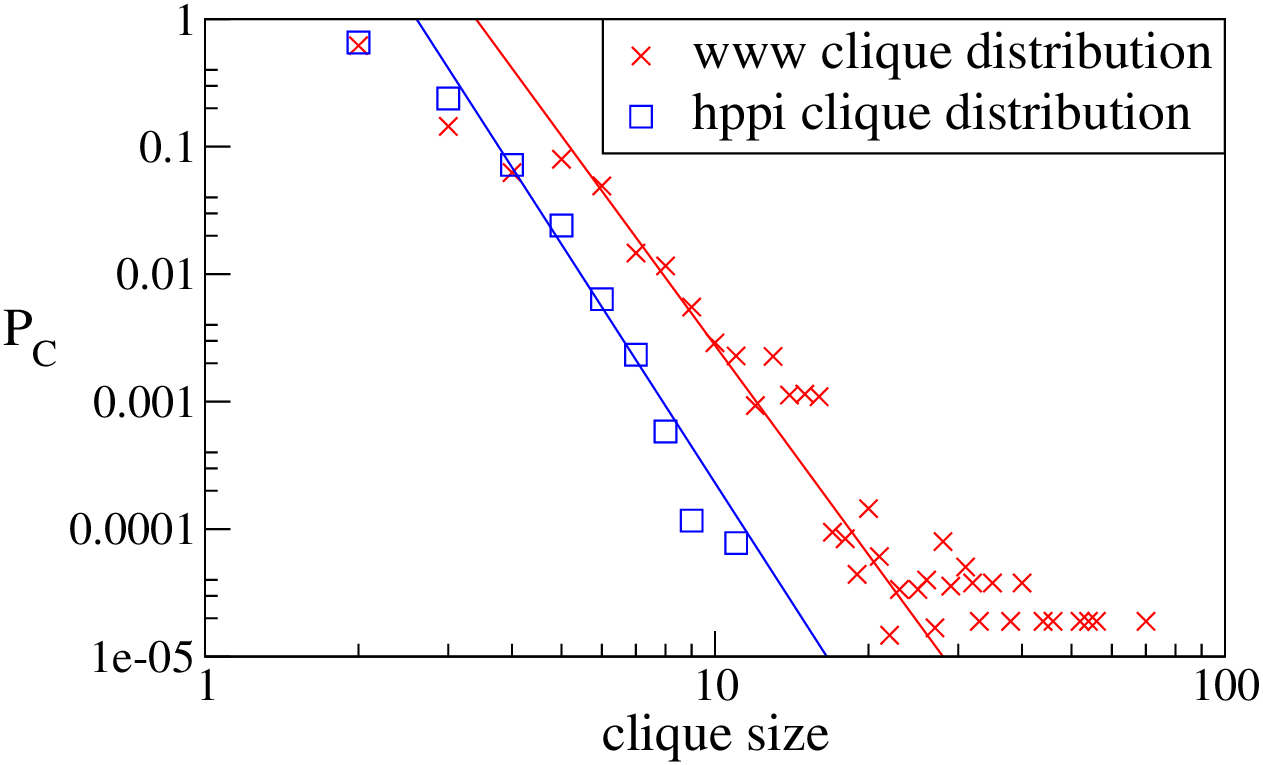}}
    \mbox{\includegraphics[width=0.45\textwidth] {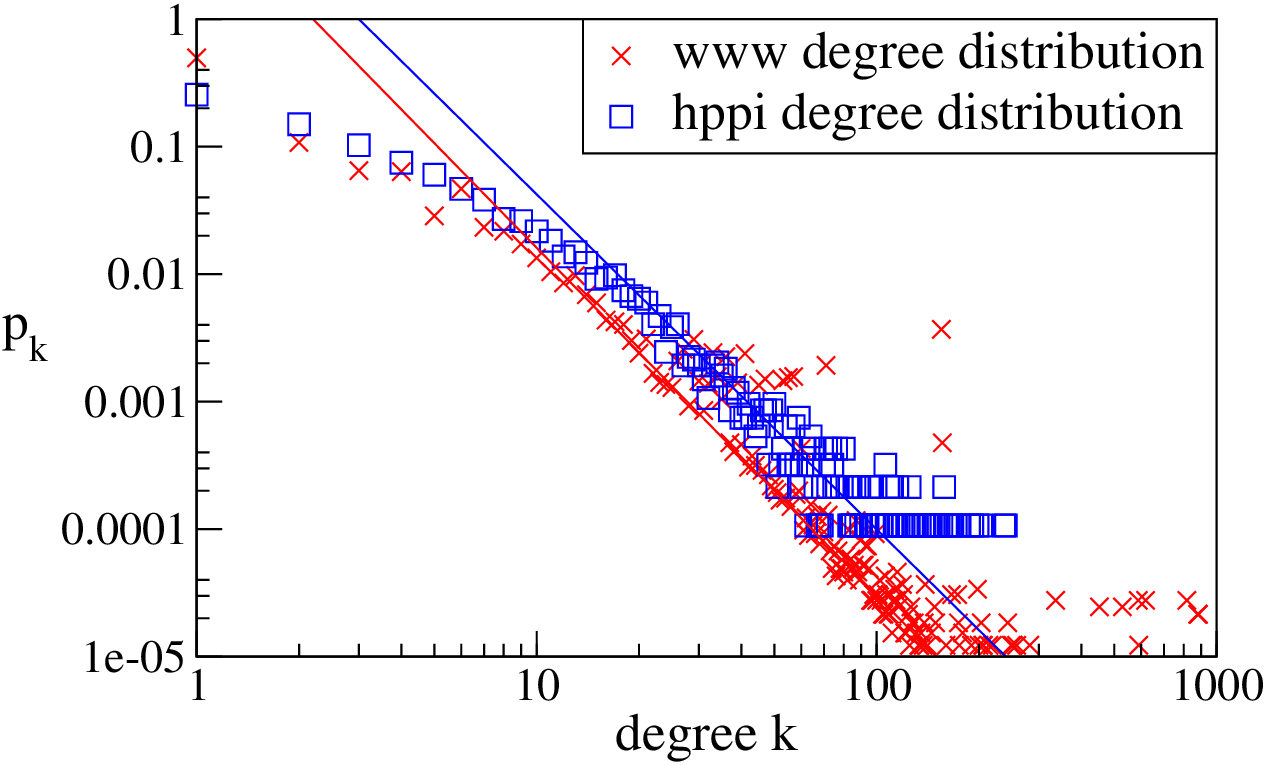}}
  }
\caption{\label{cliques} Left figure: Clique 
distribution $P_C(S)$
of the WWW data set \cite{Barabasi99} and Human Protein Protein 
Interaction Database (HPPI) \cite{Mathivanan06}. 
The distributions have the
exponents $\alpha_{www} =-5.5$, 
$\alpha_{hppi} =-6.2$.
The statistical properties are given in Table 
\ref{table_hppi_www} \newline
Right figure: Degree distribution $p_k$ of the
same data shown in left figure. Continuous 
lines show the respective slope of $m_{www}=-2.8$, $m_{hppi}=-2.5$.
The statistical properties are given in Table \ref{table_hppi_www} }
\label{www_data}
  \end{figure*}

Fig.\ \ref{cliques} shows the clique and the degree 
distributions of the respective original graphs, 
with their corresponding statistical properties given
in the Table \ref{table_hppi_www}.
We note that the protein-interaction graph contains
cliques of up-to ten sites, where a typical clique-size
is slightly larger in the WWW-net. The scaling of the
degree distribution $p_k$ is clearly observable
for the WWW-net, but only indicative for the 
protein-interaction networks, due to the limited
number of vertices it contains.


In Table \ref{table_hppi_www} we have
also included the properties of the graphs
generated by the hierarchical algorithm
using preferential attachment.
The main difference between the generated 
networks analyzed in Table \ref{table_hppi_www}
and those previously discussed, is the fact that 
they are not averaged over an ensemble of
realizations of a clique distribution.
The reason is, that the exact experimental
clique distributions for the protein-interaction network
and for the WWW-network have been taken as an
input for the hierarchical algorithm, which is
per construction conserved with respect 
to the clique distribution.

Next we note two caveats with respect to
the protein interaction graph.
Firstly, it is not complete,
being updated continuously as
new experimental results become available
\cite{Mathivanan06}. Secondly, the
protein-interaction network contains
unconnected subsets of vertices. 
The largest component does not encompass 
the entire graph but 8972 sites out of
a total of 9362 vertices. We have
used this largest component for the data
analysis.

While analyzing the data presented in 
Table \ref{table_hppi_www} we note
substantial differences between the
properties of the real-world graphs
with respect to the one generated by
the hierarchical clique-conserving 
algorithm. These differences involve
essentially all key statistical quantities,
such as the total number of
vertices, the average degree, the network
diameter and the large-k falloff of 
degree distribution.

This leaves us with two possible conclusions,
the first being that the clique distribution 
$P_C(S)$ is probably not a good quantity for
the purpose of characterizing a given graph,
at least in the two examples considered here.
The second is the possibility that an 
altogether different clique-conserving 
algorithm may be needed for the clique 
distribution to be used as a characterizing
quantity.

The data presented in Table \ref{table_hppi_www}
was generated using the hierarchical algorithm with
preferential attachment, however, as discussed above
(see Fig.\ \ref{com_rnd_prefAtt}), the difference
between random and preferential attachment is
actually quite small for clique distributions
having a fat tail.

\begin{table}[bt]
\caption{Statistical properties of a HPPI
and of a WWW graph. $C$ is the clustering 
coefficient, $\ell$ the average path
length, $\langle\kappa\rangle$ the
average degree, $D$ the diameter, $d$ the
link density, $N$ the total number
of vertices, $m$ the slope measured 
for $k\in[10,44]$ for the real data 
($k\in[10,20]$ for the generated graph and 
$k\in[10,100]$ for generated WWW data).}
\begin{tabular}{c|ccccccccccc}
\hline \hline
data  & source &  $C$   & $\ell$ & $D$ & $\langle \kappa \rangle$ & $d$ & N & $m$\\
\hline
      & real   & 0.11 & 4.3    & 14  & 7.8  & 0.00085  & 8972   & -2.5\\ 
\raisebox{1.8ex}[-1.5ex]{HPPI} 
      & gener. & 0.20 & 3.8    & 11  & 3.0  & 0.00016  & 25747  & -3.5 \\
\hline
      & real   & 0.23 & 7.2    & 46  & 3.0  & 0.000009 & 325729 & -2.8 \\
\raisebox{1.8ex}[-1.5ex]{WWW} 
      & gener. & 0.27 &  3.751 & 12  & 4.1  & 0.0000086 & 475588     & -3.6 \\
\hline \hline
\end{tabular}
\label{table_hppi_www}
\end{table}

\section{Discussion}
In this paper we presented an algorithm,
the hierarchical algorithm, by which one
can generate graphs having a pre-determined
distribution of cliques, {\it viz} of
fully connected subgraphs. We have studied,
in a first step,
the degree distribution of the resulting
networks for scale-free clique distribution
as a function of the scaling exponent.

In a second step we used two selected
real-world graphs, a protein-interaction
network and a WWW-network, and examined the
relation between their degree and
clique distributions relative to those
of graphs generated via the hierarchical
algorithm having the same respective
clique distribution. We find no good
agreement, and this leads us to the conclusion
that either the clique distribution is
insufficient for a in-depth characterization
of real-world networks or that the
hierarchical algorithms need further development.


\end{document}